\documentclass[twocolumn,showpacs,preprintnumbers,amsmath,amssymb]{revtex4}
\usepackage{graphicx}
\usepackage{dcolumn}
\usepackage{bm}
\usepackage{sidecap}
\usepackage{epstopdf}

\begin{document}

%\preprint{APS/123-QED}

%%%%%%%%%%%%%%%%%%%%%%%%%%%%%%%%%%%%%%%%%%%%%%%%%%%%%%%%%%%%%%%%%%%%%

\title{Experimental test of quantum contextuality in neutron interferometry}

%%%%%%%%%%%%%%%%%%%%%%%%%%%%%%%%%%%%%%%%%%%%%%%%%%%%%%%%%%%%%%%%%%%%%

\author{H. Bartosik$^1$}
\author{J. Klepp$^{1}$}
\author{C. Schmitzer$^1$}
\author{S. Sponar$^1$}
\author{A. Cabello$^2$}
\author{H. Rauch$^{1,3}$}
\author{Y. Hasegawa$^{1}$}
\affiliation{%
$^1$Atominstitut der \"{O}sterreichischen Universit\"{a}ten, Stadionallee 2, A-1020 Wien, Austria\\
$^2$Departamento de F\'{\i}sica Aplicada II, Universidad de Sevilla, E-41012 Sevilla, Spain\\
$^3$Institut Laue-Langevin, Boite Postale 156, F-38042 Grenoble Cedex 9, France
}

%%%%%%%%%%%%%%%%%%%%%%%%%%%%%%%%%%%%%%%%%%%%%%%%%%%%%%%%%%%%%%%%%%%%%

\date{\today}

%%%%%%%%%%%%%%%%%%%%%%%%%%%%%%%%%%%%%%%%%%%%%%%%%%%%%%%%%%%%%%%%%%%%%

\begin{abstract}
We performed an experimental test of the Kochen-Specker theorem
based on an inequality derived from the Peres-Mermin proof, using
spin-path (momentum) entanglement in a single neutron system.
Following the strategy proposed by Cabello \textit{et al.}
[Phys.~Rev.~Lett.~\textbf{100}, 130404 (2008)], a Bell-like state
was generated and three expectation values were determined. The
observed violation $2.291 \pm 0.008 \not \leq 1$ clearly shows that
quantum mechanical predictions cannot be reproduced by noncontextual
hidden variables theories.
\end{abstract}

%%%%%%%%%%%%%%%%%%%%%%%%%%%%%%%%%%%%%%%%%%%%%%%%%%%%%%%%%%%%%%%%%%%%%

\pacs{03.65.Ud, 03.75.Dg, 07.60.Ly, 42.50.Xa}
% PACS, the Physics and Astronomy

\maketitle

%First version: unknown
%This version: 23 July 2009 after PRL published version PRL 103, 040403 (2009).

%%%%%%%%%%%%%%%%%%%%%%%%%%%%%%%%%%%%%%%%%%%%%%%%%%%%%%%%%%%%%%%%%%%%%

There are two powerful arguments against the possibility of
extending quantum mechanics (QM) into a more fundamental theory
yielding a deterministic description of nature. One is the
experimental violation of Bell inequalities \cite{Bell:1964rw,
PhysRevLett.49.91, PhysRevLett.81.5039, Rowe:2001qy, scheidl-2008,
Hasegawa:2003ty}, which discards local hidden-variable theories as a
possible extension to QM. The other is the Kochen-Specker (KS)
theorem \cite{ISI:A19679627500004}, which stresses the
incompatibility of QM with a larger class of hidden-variable
theories, known as noncontextual hidden-variable theories (NCHVTs).
By definition, NCHVTs assume that the result of a measurement of an
observable is predetermined and independent of a suitable (previous
or simultaneous) measurement of any other compatible
(i.e.,~comeasurable) observable. While the original proof of the KS
theorem is rather complicated, simplified versions have been
proposed by Peres \cite{Peres90} and Mermin
\cite{PhysRevLett.65.3373, RevModPhys.65.803}. These proofs can be
converted into experimentally testable inequalities, valid for any
NCHVT, but violated by QM \cite{cabello:130404, Cabello08}.

Since the first observation of neutron self-interference
\cite{Rauch1974369}, neutron optical experiments have been serving
as an established method for investigating the foundations of
quantum mechanics. In particular, neutron interferometry allows the
observation of quantum mechanical phenomena on a macroscopic scale
\cite{Rauch:2000hl}. Studies on entanglement between 2 degrees of
freedom of single neutrons confirmed the violation of Bell-like
inequalities \cite{Hasegawa:2003ty}. A complete tomographic
reconstruction of density matrices was performed
\cite{hasegawa:052108}. Recent developments on the coherent
manipulation of the energy degree of freedom of single neutrons
\cite{sponar:061604} provide the basis for the generation of triply
entangled quantum states, where the peculiarity of a
Greenberger-Horne-Zeilinger-like state was exhibited
\cite{Hasegawa:2009qy}. In addition to the interferometric scheme,
the nonadditivity of the mixed state phase was demonstrated in
neutron polarimetry \cite{klepp:150404}. Neutrons in the
ultralow-energy regime, i.e., ultracold neutrons (UCNs), can be
stored for several minutes, which allows for novel explorations: the
stability of the Berry phase was studied by tuning the evolution
time during the storage \cite{filipp:030404}. At a stage of
experimental tests of quantum contextuality, the spin-path
(momentum) entanglement in single neutrons allowed for demonstrating
Kochen-Specker-like phenomena \cite{2006PhRvL..97w0401H}. Further
theoretical analysis revealed a more advanced scheme and an
experiment with neutron interferometry was proposed
\cite{cabello:130404}. In this Letter we report on an improved
experimental test of the KS theorem using single neutrons entangled
in 2 degrees of freedom.

%%%%%%%%%%%%%%%%%%%%%%%%%%%%%%%%%%%%%%%%%%%%%%%%%%%%%%%%%%%%%%%%%%%%%

For the proof of the Kochen-Specker theorem, we consider single
neutrons prepared in a maximally entangled Bell-like state described
by the wave function
\begin{equation}
| \Psi \rangle = \frac{1}{\sqrt{2}} (\mid \downarrow \rangle \otimes
| I \rangle - \mid \uparrow \rangle \otimes | II
\rangle),\label{prepstate}
\end{equation}
where $\mid \uparrow \rangle$ and $\mid \downarrow \rangle$ denote
the up-spin and down-spin eigenstates, and $| I \rangle$ and $| II
\rangle$ the two beam paths in a neutron interferometer. We define
Pauli-type operators for the spin and path degree of freedom,
e.g.,~$\sigma ^{s} _{x} = \mid \uparrow \rangle \langle \downarrow
\mid + \mid \downarrow \rangle \langle \uparrow \mid$ and $\sigma
^{p} _{x} = | I \rangle \langle II | + | II \rangle \langle I |$,
where $s$ stands for spin and $p$ for path. The proof is based on
the six observables $\sigma_{x}^{s}$, $\sigma_{x}^{p}$,
$\sigma_{y}^{s}$, $\sigma_{y}^{p}$, $\sigma_{x}^{s}\sigma_{y}^{p}$,
and $\sigma_{y}^{s}\sigma_{x}^{p}$, and the following five quantum
mechanical predictions for the state $|\Psi\rangle$,
\begin{subequations}
\begin{align}
\sigma ^{s} _{x} \cdot \sigma ^{p} _{x} \:| \Psi \rangle &= -| \Psi
\rangle, \label{qmpredictionsa}
\\
\sigma ^{s} _{y} \cdot \sigma ^{p} _{y} \:| \Psi \rangle &= -| \Psi
\rangle, \label{qmpredictionsb}
\\
\sigma ^{s} _{x} \sigma ^{p} _{y} \cdot \sigma ^{s} _{x} \cdot
\sigma ^{p} _{y} \:| \Psi \rangle & = +| \Psi \rangle,
\label{qmpredictionsc}
\\
\sigma ^{s} _{y} \sigma ^{p} _{x} \cdot \sigma ^{s} _{y} \cdot
\sigma ^{p} _{x} \:| \Psi \rangle &= +| \Psi\rangle,
\label{qmpredictionsd}
\\
\sigma ^{s} _{x} \sigma ^{p} _{y} \cdot \sigma ^{s} _{y} \sigma ^{p}
_{x} \:| \Psi \rangle &= -| \Psi \rangle. \label{qmpredictionse}
\end{align}
\end{subequations}
Reproducing these predictions in the framework of NCHVTs requires
assigning predetermined measurement results to each of the six
observables [note that we introduced ($\cdot$) to separate operators
which, in NCHVTs, correspond to observables with predetermined
measurement results]. The inconsistency arising in any attempt to
ascribe the predefined values $-1$ or $+1$ to each and every of the
six observables can be easily seen by multiplying
Eqs.~(\ref{qmpredictionsa})--(\ref{qmpredictionse}). Since each
observable appears twice, the left hand sides give $+1$ while the
product of the right hand sides is $-1$.

An ideal experiment for verifying this contradiction would be to
confirm each of the five predictions of QM,
Eqs.~(\ref{qmpredictionsa})--(\ref{qmpredictionse}). However, it is
not possible to obtain perfect correlations in a real experiment. An
experimentally testable inequality can be derived from the linear
combination of the five expectation values with the respective
quantum mechanical predictions as linear coefficients. It can be
shown that in any NCHVT
\begin{eqnarray}
-\langle \sigma ^{s} _{x} \cdot \sigma ^{p} _{x} \rangle - \langle
\sigma ^{s} _{y} \cdot \sigma ^{p} _{y} \rangle + \langle \sigma ^{s} _{x}
\sigma ^{p} _{y} \cdot \sigma ^{s} _{x} \cdot \sigma ^{p} _{y} \rangle
\hfill\qquad\qquad\:\nonumber\\
+ \:\langle \sigma ^{s} _{y} \sigma ^{p} _{x} \cdot \sigma ^{s} _{y}
\cdot \sigma ^{p} _{x} \rangle - \langle \sigma ^{s} _{x} \sigma
^{p} _{y} \cdot \sigma ^{s} _{y} \sigma ^{p} _{x} \rangle \leq 3,
\label{inequality3}
\end{eqnarray}
while the prediction of QM is 5. In order to test this inequality
one needs to perform five experiments according to the five
different experimental contexts represented by
Eqs.~(\ref{qmpredictionsa})--(\ref{qmpredictionse}). It is important
to note here that (since we would like to test quantum
contextuality) the six measurement apparatuses used for measuring
the six observables must be the same irrespective of the
experimental context in which they appear.

As already pointed out in the previous Letter \cite{cabello:130404},
the five experiments of
Eqs.~(\ref{qmpredictionsa})--(\ref{qmpredictionse}) contribute in
different ways to the proof. While Eqs.~(\ref{qmpredictionsa}),
(\ref{qmpredictionsb}), and (\ref{qmpredictionse}) represent
state-dependent predictions relying on the specific properties of
the state $|\Psi\rangle$, Eqs.~(\ref{qmpredictionsc}) and
(\ref{qmpredictionsd}) are state-independent predictions which hold
in any NCHVT. In other words, in any NCHVT, $\langle \sigma ^{s}
_{x} \sigma ^{p} _{y} \cdot \sigma ^{s} _{x} \cdot \sigma ^{p} _{y}
\rangle = 1$ and $\langle \sigma ^{s} _{y} \sigma ^{p} _{x} \cdot
\sigma ^{s} _{y} \cdot \sigma ^{p} _{x} \rangle =1$. Therefore, any
NCHVT must satisfy not only inequality~(\ref{inequality3}), but also
the following inequality:
\begin{equation}
-\langle \sigma ^{s} _{x} \cdot \sigma ^{p} _{x} \rangle - \langle
\sigma ^{s} _{y} \cdot \sigma ^{p} _{y} \rangle - \langle \sigma
^{s} _{x} \sigma ^{p} _{y} \cdot \sigma ^{s} _{y} \sigma ^{p} _{x}
\rangle \leq 1. \label{inequtbtested}
\end{equation}
A violation of inequality~(\ref{inequtbtested}) reveals quantum
contextuality as long as the measurements of the six observables
involved in~(\ref{inequtbtested}) are performed in such a way that
it would be possible to determine
 also $\langle \sigma ^{s} _{x} \sigma
^{p} _{y} \cdot \sigma ^{s} _{x} \cdot \sigma ^{p} _{y} \rangle$ and
$\langle \sigma ^{s} _{y} \sigma ^{p} _{x} \cdot \sigma ^{s} _{y}
\cdot \sigma ^{p} _{x} \rangle$, at least in principle. In this
Letter, we report on an experimental test of
inequality~(\ref{inequtbtested}), including a prescription of how
the measurement apparatuses used in our experiments can be combined
for realizing a test of inequality~(\ref{inequality3}).

%%%%%%%%%%%%%%%%%%%%%%%%%%%%%%%%%%%%%%%%%%%%%%%%%%%%%%%%%%%%%%%%%%%%%
% Experiment
%%%%%%%%%%%%%%%%%%%%%%%%%%%%%%%%%%%%%%%%%%%%%%%%%%%%%%%%%%%%%%%%%%%%%

\begin{figure}[t]
\centering
\includegraphics[width=.44\textwidth]{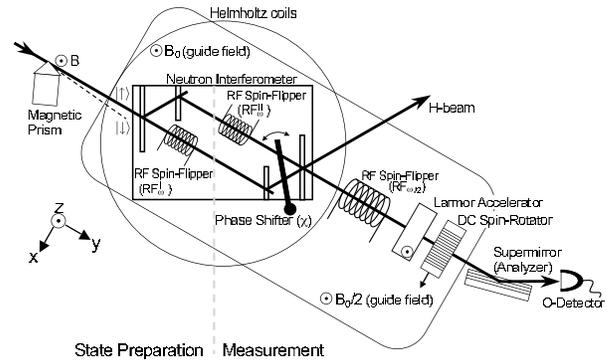}
\caption{Experimental setup: The maximally entangled Bell state
$|\Psi \rangle$ is generated in the first half of a skew-symmetric
interferometer. The second half of the interferometer together with
a phase shifter serves as a path measurement apparatus. A spin
analysis system in the O beam allows for the selection of neutrons
with certain spin properties. The spin flipper in path $|II\rangle$
is required for the measurement of the product observable $\langle
\sigma_{x}^{s}\sigma_{y}^{p}\cdot\sigma_{y}^{s}\sigma_{x}^{p}\rangle$.}
\label{Fig1}
\end{figure}

The experiment was carried out at the perfect-crystal neutron optics
beamline S18 at the high flux reactor of the Institute Laue-Langevin
(ILL). A triple Laue interferometer setup (see Fig.~\ref{Fig1})
similar to previous neutron interferometric experiments
\cite{sponar:061604} was used. By means of a Si perfect-crystal
monochromator, a neutron beam with a mean wavelength of $\lambda
_{0}=$ 1.92 \AA~($\Delta\lambda/\lambda_{0} \sim0.02$) is selected.
The incident beam is confined to a beam cross section of $5\times5$
mm$^{2}$ and polarized in the vertical direction using the spin
dependent birefringence in two sequential magnetic prisms. Because
of the angular separation of the two sub-beams, only up-spin
neutrons $(\mid \uparrow \rangle)$ meet the Bragg condition at the
first interferometer plate and are split coherently into the two
spatially separated paths, $| I \rangle$ and $| II \rangle$.
Together with a radio-frequency (RF) spin flipper in path $| I
\rangle$, denoted as RF$_{\omega}^I$, the first half of the
interferometer is used for the generation of the maximally entangled
Bell-like state $|\Psi \rangle$ [Eq.~(\ref{prepstate})]. A
parallel-sided Si plate serves as a phase shifter for the path
degree of freedom prior to the coherent recombination of the two
paths at the third interferometer plate. Only neutrons emerging the
interferometer in the forward direction (O beam) are used for the
measurements. As explained below, our experiment requires a second
RF flipper in the interferometer (RF$_{\omega}^{I\!I}$) and another
RF flipper in the O beam operated at half frequency (RF$_{\omega
/2}$). Two pairs of water-cooled Helmholtz coils create a fairly
uniform magnetic guide field $ B_{0} \hat z$ of $B_{0} \simeq 20$ G
and $B_{0} / 2 \simeq 10$ G in the region of the interferometer and
alongside the O beam, respectively. A spin analyzing supermirror
(transmitting up-spin neutrons only) in combination with additional
direct current (DC) spin rotators enable arbitrary measurements of
the spin degree of freedom in the O beam: neutrons with the selected
spin properties are counted in the subsequent O detector (efficiency
$>99\%$).

%%%%%%%%%%%%%%%%%%%%%%%%%%%%%%%%%%%%%%%%%%%%%%%%%%%%%%%%%%%%%%%%%%%%%
% Figure 1
%%%%%%%%%%%%%%%%%%%%%%%%%%%%%%%%%%%%%%%%%%%%%%%%%%%%%%%%%%%%%%%%%%%%%

%%%%%%%%%%%%%%%%%%%%%%%%%%%%%%%%%%%%%%%%%%%%%%%%%%%%%%%%%%%%%%%%%%%%%

\begin{figure}[b]
  \centering
  \includegraphics [width=0.4\textwidth]{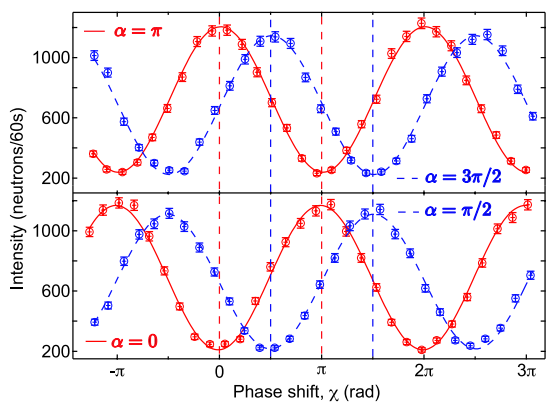}
  \caption{
Typical intensity modulations obtained by varying the phase $\chi$
for the path subspace. The contrast of the sinusoidal oscillations
is above $67\%$. The parameter $\alpha$ represents the direction of
the spin analysis. In particular, the settings $\alpha=0, \pi$
$[\alpha=\tfrac{\pi}{2},\tfrac{3 \pi}{2}]$ were used for measuring
$\sigma^s_x$ $[\sigma^s_y]$. The expectation values, $E(0,0)$ $[E(
\frac{\pi}{2}, \frac{\pi}{2})]$, are determined from the intensities
on the dashed lines, $\chi=0, \pi$ $[\chi=\tfrac{\pi}{2},\tfrac{3
\pi}{2}]$.
    }
\label{fig:2}
\end{figure}

The first term in inequality (\ref{inequtbtested}) requires the
measurement of $\sigma ^{s} _{x}$ together with $\sigma ^{p} _{x}$.
Here, RF$_{\omega /2}$ in the O beam is needed for compensating the
energy difference due to the spin flip at RF$_{\omega}^I$
\cite{sponar:061604}, while the second RF flipper in the
interferometer, RF$_{\omega}^{I\!I}$, is turned off. For measuring
the path observable, i.e., $\sigma ^{p} _{x}$, the phase shifter is
adjusted to induce a relative phase ($\chi$) between the two paths
$| I \rangle$ and $| II \rangle$. Settings of $ \chi = 0 $ and $
\chi = \pi $ in the path state $|\Psi(\chi)\rangle_{p} =
\frac{1}{\sqrt{2}} (| I \rangle + e^{i \chi} | II \rangle)$
correspond to the projections to $|+x\rangle_{p}$ and
$|-x\rangle_{p}$, the two eigenstates of $\sigma ^{p} _{x}$,
respectively. The spin analysis in the $x$-$y$ plane is accomplished
by the combination of the Larmor accelerator DC coil inducing a
Larmor phase $\alpha$, the DC spin rotator tuned to a $\pi / 2$
rotation and the analyzing supermirror. This configuration allows
for the selection of neutrons in the spin state $|\Psi (\alpha)
\rangle_{s} = \frac{1}{\sqrt{2}} (\mid \uparrow \rangle + e^{i
\alpha}\mid \downarrow \rangle )$. Spin analysis in arbitrary
directions of the $x$-$y$ plane can be realized by adequately
adjusting the Larmor phase $\alpha$ between $0$ and $2\pi$. For
example, the projections to $|+x\rangle_{s}$ and $|-x\rangle_{s}$,
the two eigenstates of $\sigma ^{s} _{x}$, correspond to $\alpha =
0$ and $\alpha = \pi$, respectively. The experimental setup for the
second term in inequality (\ref{inequtbtested}) is identical with
the one for the first term, but the measurement of $\sigma ^{s}
_{y}$ together with $\sigma ^{p} _{y}$ is achieved with settings of
$\chi = \frac{\pi}{2}, \frac{3\pi}{2}$ and $\alpha = \frac{\pi}{2} ,
\frac{3\pi}{2}$. Typical intensity oscillations for the successive
measurement of the path and the spin component are shown in
Fig.~\ref{fig:2}. Clear sinusoidal dependence of the intensity on
the relative phase shift $\chi$ is observed. The corresponding
expectation values are then derived from the relation
\begin{equation}
E(\alpha, \chi) = \tfrac{N(\alpha, \chi) + N(\alpha + \pi, \chi +
\pi) - N(\alpha + \pi, \chi) - N(\alpha, \chi + \pi)}{N(\alpha,
\chi) + N(\alpha + \pi, \chi + \pi) + N(\alpha + \pi, \chi) +
N(\alpha, \chi + \pi)}\label{equ:expvalue},
\end{equation}
where $N(\alpha, \chi)$ denotes the neutron count rate at the joint
projection to the spin state $|\Psi (\alpha)\rangle_{s}$ and the
path state $|\Psi (\chi)\rangle_{p}$. The required count rates at
appropriate settings of $\alpha$ and $\chi$ (indicated by the
vertical dashed lines in Fig.~\ref{fig:2}) are determined from least
squares fits. Each measurement was carried out 3 times in order to
reduce statistical errors. All errors of the fit parameters and the
experimentally unavoidable phase drifts are included in the error
estimation. In this way, we obtain the expectation values $\langle
\sigma ^{s} _{x} \cdot \sigma ^{p} _{x} \rangle \equiv E(0, 0) =
-0.679 \pm 0.005$ and $\langle \sigma ^{s} _{y} \cdot \sigma ^{p}
_{y} \rangle \equiv E(\frac{\pi}{2}, \frac{\pi}{2})= -0.682 \pm
0.005$.

%%%%%%%%%%%%%%%%%%%%%%%%%%%%%%%%%%%%%%%%%%%%%%%%%%%%%%%%%%%%%%%%%%%%%
% Figure 2
%%%%%%%%%%%%%%%%%%%%%%%%%%%%%%%%%%%%%%%%%%%%%%%%%%%%%%%%%%%%%%%%%%%%%

%%%%%%%%%%%%%%%%%%%%%%%%%%%%%%%%%%%%%%%%%%%%%%%%%%%%%%%%%%%%%%%%%%%%%

The third term in inequality (\ref{inequtbtested}) requires the
measurement of $\sigma ^{s} _{x} \sigma ^{p} _{y}$ together with
$\sigma ^{s} _{y} \sigma ^{p} _{x}$. Measuring the product of these
two observables simultaneously implies the discrimination of the
four possible outcomes $(\sigma ^{s} _{x} \sigma ^{p} _{y} , \sigma
^{s} _{y} \sigma ^{p} _{x}) =
\left\{(+1,+1),(-1,-1),(+1,-1),(-1,+1)\right\},$ which is equivalent
to a complete Bell-state discrimination \cite{2005PhRvL..95x0406Y,
barbieri:042317}. The two operators $\sigma ^{s} _{x} \sigma ^{p}
_{y}$ and $\sigma ^{s} _{y} \sigma ^{p} _{x}$ have the four common
Bell-like eigenstates
\begin{subequations}
\begin{align}
| \varphi _{\pm} \rangle &= \tfrac{1}{\sqrt{2}} ( \mid \downarrow
\rangle \otimes | I \rangle \pm i \mid \uparrow \rangle \otimes | II
\rangle),\label{def:varphi+-}
\\
| \phi _{\pm} \rangle &= \tfrac{1}{\sqrt{2}} ( \mid \uparrow \rangle
\otimes | I \rangle \pm i \mid \downarrow \rangle \otimes | II
\rangle),\label{def:phi+-}
\end{align}
\end{subequations}
with the corresponding eigenvalue equations
\begin{subequations}
\begin{align}
\sigma ^{s} _{x} \sigma ^{p} _{y} ~| \varphi _{\pm} \rangle &= \pm |
\varphi _{\pm} \rangle, \qquad &\sigma ^{s} _{y} \sigma ^{p} _{x} ~|
\varphi _{\pm} \rangle &= \mp | \varphi _{\pm} \rangle,
\label{xsyp-ysxpEW1}
\\
\sigma ^{s} _{x} \sigma ^{p} _{y} ~| \phi _{\pm} \rangle &= \pm |
\phi _{\pm} \rangle, \qquad & \sigma ^{s} _{y} \sigma ^{p} _{x} ~|
\phi _{\pm} \rangle &= \pm | \phi _{\pm} \rangle.
\label{ysxp-xsypEW2}
\end{align}
\end{subequations}
It follows that the outcome $-1$ for the product measurement of
$\sigma ^{s} _{x} \sigma ^{p} _{y}$ and $\sigma ^{s} _{y} \sigma
^{p} _{x}$ is obtained for $| \varphi _{\pm} \rangle$, while the
states $| \phi _{\pm} \rangle$ yield the result $+1$. In practice,
this Bell-state discrimination is accomplished by the following
setup: RF$_{\omega}^{I\!I}$ is tuned to flip the spin in path $II$,
i.e.,~transforming the state $| \Psi \rangle \rightarrow
\frac{1}{\sqrt{2}} (\mid \downarrow \rangle \otimes | I \rangle -
\mid \downarrow \rangle \otimes | II \rangle)$. Note that
RF$_{\omega /2}$ (used for compensating an energy difference between
the two subbeams) is not needed for this measurement because the
energy of the two sub-beams in the interferometer is the same after
the spin flip in each path.

%%%%%%%%%%%%%%%%%%%%%%%%%%%%%%%%%%%%%%%%%%%%%%%%%%%%%%%%%%%%%%%%%%%%%
% Figure 3
%%%%%%%%%%%%%%%%%%%%%%%%%%%%%%%%%%%%%%%%%%%%%%%%%%%%%%%%%%%%%%%%%%%%%

\begin{figure}[b]
\begin{center}
\includegraphics[width=0.4\textwidth]{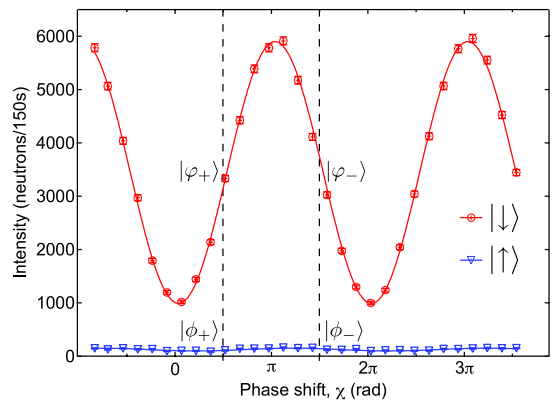}
\caption{ Typical intensity modulations obtained by varying the
phase $\chi$ in the path subspace. The two curves were recorded with
opposite settings of the spin-analysis system selecting $\mid
\downarrow \rangle$- and $\mid \uparrow \rangle$- components,
respectively. Intensities on the dashed lines
($\chi=\tfrac{\pi}{2},\tfrac{3 \pi}{2}$) are used for the evaluation
of the expectation value $\langle \sigma ^{s} _{x} \sigma ^{p} _{y}
\cdot \sigma ^{s} _{y} \sigma ^{p} _{x} \rangle$. } \label{fig:3}
\end{center}
\end{figure}

%%%%%%%%%%%%%%%%%%%%%%%%%%%%%%%%%%%%%%%%%%%%%%%%%%%%%%%%%%%%%%%%%%%%%

When the DC spin rotator in the O beam is adjusted to induce a $\pi$
flip, only $\mid\downarrow \rangle$-spin components reach the
detector. Inducing a relative phase $\chi$ between the two beam
paths in the interferometer allows then for projections to the state
$| \varphi (\chi)\rangle = \frac{1}{\sqrt{2}}(\mid \downarrow
\rangle \otimes | I \rangle + e^{i \chi} \mid \uparrow \rangle
\otimes | II \rangle)$. According to Eq.~(\ref{def:varphi+-}), phase
settings of $\chi = \pm \frac{\pi}{2}$ correspond to the measurement
of $| \varphi _{\pm} \rangle$. On the other hand, $\mid\uparrow
\rangle$-spin analysis is achieved by switching the DC spin rotator
off, where neutrons in the state $| \phi (\chi)\rangle =
\frac{1}{\sqrt{2}}(\mid \uparrow \rangle \otimes | I \rangle + e^{i
\chi} \mid \downarrow \rangle \otimes | II \rangle)$ can be
selected. Comparing with Eq.~(\ref{def:phi+-}), projections to
$|\phi _{\pm} \rangle$ are obtained with the phase shifter settings
$\chi=\pm \frac{\pi}{2}$. Typical intensity modulations for the two
opposite settings of the spin analysis are shown in
Fig.~\ref{fig:3}. Clear sinusoidal intensity oscillation is observed
for analyzing $\mid \downarrow \rangle$ components, whereas the
intensities with $\mid \uparrow \rangle$ spin analysis are marginal.
The two relevant settings of the phase shifter, $\chi = \pm
\frac{\pi}{2}$, are indicated by vertical dashed lines. The observed
intensities reflect the quantum mechanical predictions for the
measurement of the four Bell-like states given by the expectation
values $\langle \Psi | \varphi _{\pm} \rangle \langle \varphi _{\pm}
| \Psi \rangle = \frac{1}{2}$ and $\langle \Psi | \phi _{\pm}
\rangle \langle \phi _{\pm} | \Psi \rangle = 0$. Because of
experimental imperfections, e.g., slightly less than 100\% incident
polarization and efficiencies of the spin flips, small contributions
from the $| \phi _{\pm} \rangle$ components were found. Note that
the setting of $\chi = \pi$ in the down-spin analysis yields the
projection to $| \Psi\rangle = \tfrac{e^{i\pi/4}}{\sqrt{2}}(
|\varphi _{+} \rangle - i |\varphi _{-} \rangle )$. The intensity
maximum located at this setting clearly proves the correct
preparation of the state $ |\Psi \rangle$. The expectation value
$\langle \sigma ^{s} _{x} \sigma ^{p} _{y} \cdot \sigma ^{s} _{y}
\sigma ^{p} _{x} \rangle$ is derived from the relation
\begin{equation}
E' = \tfrac{N'(\phi _{+}) + N'(\phi _{-}) - N'(\varphi _{+}) - N'(
\varphi _{-})}{N'(\phi _{+})+ N'(\phi _{-}) + N'(\varphi _{+}) +
N'(\varphi _{-})}\label{equ:expvaluebelldiscr},
\end{equation}
where $N'(\ldots)$ denotes the neutron count rate at the indicated
projections. As before, least square fits were applied to deduce the
count rates at the four projections. Because of thermal disturbances
from the RF spin flippers in the interferometer, systematic shifts
of up to $9^{\circ}$ of the measured oscillations were observed.
Including all experimental errors in the error estimation, we
determine the expectation value $\langle \sigma ^{s} _{x} \sigma
^{p} _{y} \cdot \sigma ^{s} _{y} \sigma ^{p} _{x} \rangle \equiv E'
= -0.93 \pm 0.003$.

With the three experimentally derived expectation values we can
finally test inequality (\ref{inequtbtested}). We obtain $-\langle
\sigma ^{s} _{x} \cdot \sigma ^{p} _{x} \rangle -\langle \sigma ^{s}
_{y} \cdot \sigma ^{p} _{y} \rangle - \langle \sigma ^{s} _{x}
\sigma ^{p} _{y} \cdot \sigma ^{s} _{y} \sigma ^{p} _{x} \rangle =
2.291 \pm 0.008$, which is well above the upper limit of 1 given by
the bound of NCHVTs. This result represents a violation of
inequality (\ref{inequtbtested}) by 170 standard deviations.
Moreover, the measured value is evidently closer to the quantum
mechanical prediction of 3 than to the limit of NCHVTs.

As mentioned above, we also need to provide a prescription on how we
could, at least in principle, test the experimental contexts of
Eqs.~(\ref{qmpredictionsc}) and (\ref{qmpredictionsd}). In case of
testing Eq.~(\ref{qmpredictionsc}), the two possible outcomes of a
measurement of $\sigma ^{s} _{x} \sigma ^{p} _{y}$ have to be
discriminated with the same apparatus used to measure $\langle
\sigma ^{s} _{x} \sigma ^{p} _{y} \cdot \sigma ^{s} _{y} \sigma ^{p}
_{x} \rangle$. In order to perform the consecutive measurement of
$\sigma ^{s} _{x} \cdot \sigma ^{p} _{y}$, the information of the
comeasured observable $\sigma ^{s} _{y}\sigma ^{p} _{x}$ has to be
erased. From the four output channels of the Bell-state
discrimination apparatus, $| \varphi_{+}\rangle$ and $|
\phi_{+}\rangle$ ($| \varphi_{-}\rangle$ and $| \phi_{-}\rangle$)
correspond to the result $+1$ ($-1$) for the measurement of $\sigma
^{s} _{x} \sigma ^{p} _{y}$. It can be shown, that a coherent
superposition of the $| \varphi_{+}\rangle$ and $| \phi_{+}\rangle$
($| \varphi_{-}\rangle$ and $| \phi_{-}\rangle$) components at a
beam splitter with subsequent path and spin manipulation in form of
unitary state rotations allows to preserve the information on the
observable $\sigma ^{s} _{x} \sigma ^{p} _{y}$, while erasing any
knowledge on the comeasured observable $\sigma ^{s} _{y}\sigma ^{p}
_{x}$. The resulting sub-beams are then analyzed in path and spin
degree of freedom with the same apparatus used for measuring $
\langle \sigma ^{s} _{x} \cdot \sigma ^{p} _{x} \rangle$ and $
\langle \sigma ^{s} _{y} \cdot \sigma ^{p} _{y} \rangle$. A similar
experimental setup can be used for the measurement of $\langle\sigma
^{s} _{y} \sigma ^{p} _{x} \cdot \sigma ^{s} _{y} \cdot \sigma ^{p}
_{x}\rangle$. With these configurations, the two missing
experimental contexts of Eqs.~(\ref{qmpredictionsc}) and
(\ref{qmpredictionsd}) can be tested.

Although independent manipulation of energy and spin degrees of
freedom in neutron interferometry were reported
\cite{sponar:061604,Hasegawa:2009qy}, these two Hilbert subspaces
are always coupled in the experiment performed here and only
spin-path entangled states need to be considered.

In summary, we entangled the spin and the path degrees of freedom of
single neutrons in neutron interferometry for testing an inequality
based on the Peres-Mermin proof of the Kochen-Specker theorem. The
three expectation values required for the proof were obtained in
sequential measurements. In particular, one of the expectation
values was derived from a Bell-state discrimination method. The
observed value of $2.291\pm0.008 \not\leq1$ clearly confirms the
conflict with NCHVTs.

We appreciate helpful discussions with R.\,A. Bertlmann and K.
Durstberger-Rennhofer (Vienna). This work was supported by the
Japanese Science and Technology Agency, the Austrian Fonds zur
F\"{o}rderung der Wissenschaftlichen Forschung, the MCI Project No.
FIS2008-05596, and the Junta de Andaluc\'{\i}a Excellence Project
No. P06-FQM-02243.

%%%%%%%%%%%%%%%%%%%%%%%%%%%%%%%%%%%%%%%%%%%%%%%%%%%%%%%%%%%%%%%%%%%%%

%\bibliography{D:/Neutronen/Diss/Tex/BibTex/references}
%\bibliography{D:\Neutronen\Diss\IFM Kochen-Specker&Legget Typ\KS paper\references.bib}

\end{document}